\documentclass[aps, prb, reprint]{revtex4-1}
\usepackage{graphicx}
\usepackage{color}
\begin{document}
\newcommand{\eqlabel}[1]{\label{eq:#1}}
\newcommand{\eqref}[1]{Eq.~(\ref{eq:#1})}
\newcommand{\figlabel}[1]{\label{fig:#1}}
\newcommand{\figref}[1]{Fig.~\ref{fig:#1}}
\newcommand{\myabb}[2]{\textit{#1~}(#2)}
\title{
  Numerical study of incommensurability of the spiral state on
  spin-1/2 spatially anisotropic triangular antiferromagnets using
  entanglement renormalization
}
\author{Kenji Harada}
\affiliation{Graduate School of Informatics, Kyoto University, Kyoto 606-8501, Japan}
%
\begin{abstract}
  The ground state of an S=1/2 antiferromagnetic Heisenberg model on a
  spatially anisotropic triangular lattice, which is an effective
  model of Mott insulators on a triangular layer of organic charge
  transfer salts or Cs${}_2$CuCl${}_4$, is numerically studied. We
  apply a numerical variational method by using a tensor network with
  entanglement renormalization, which improves the capability of
  describing a quantum state. Magnetic ground states are identified
  for $0.7 \le J_2/J_1 \le 1$ in the thermodynamic limit, where $J_1$
  and $J_2$ denote the inner-chain and inter-chain coupling constants,
  respectively. Except for the isotropic case ($J_1=J_2$), the
  magnetic structure is spiral with an incommensurate wave vector that
  is different from the classical one. The quantum fluctuation weakens
  the effective coupling between chains, but the magnetic order
  remains in the thermodynamic limit. In addition, the incommensurate
  wave number is in good agreement with that of the series expansion
  method.
\end{abstract}
\maketitle

%
%

\section{Introduction}
\label{sec:intro}
The physics of Mott insulators has been attracting attention since the
discovery of high-temperature superconductors. In the past decade, a
number of new Mott insulator materials on a triangular layer have been
found. For example, Cs${}_2$CuCl${}_4$\cite{{Coldea:2001bz}} and
organic charge transfer salts\cite{[{See a review article:
  }][]{Powell:2011ce}}, such as $\kappa$-(BEDT-TTF)${}_2$ X and
$\beta'$-Z[Pd(dmit)${}_2$]${}_2$, have been extensively studied using
experimental and theoretical approaches. At low temperatures in the
Mott insulator phase, these materials show various equilibrium quantum
states: an antiferromagnetic long-range ordered state, a valence bond
crystal state, and a disordered state. In particular, the disordered
behaviors in $\kappa$-(BEDT-TTF)${}_2$ Cu${}_2$
(CN)${}_3$\cite{Shimizu:2003gq, Yamashita:2008ca, Yamashita:2009dl},
EtMe${}_3$Sb[Pd(dmit)${}_2$]${}_2$\cite{Itou:2008fg,
  Yamashita:2010ie}, and Cs${}_2$CuCl${}_4$\cite{{Coldea:2001bz}} are
of great interest.

\begin{figure}[ht]
  \centering
  \includegraphics[width=0.4\textwidth]{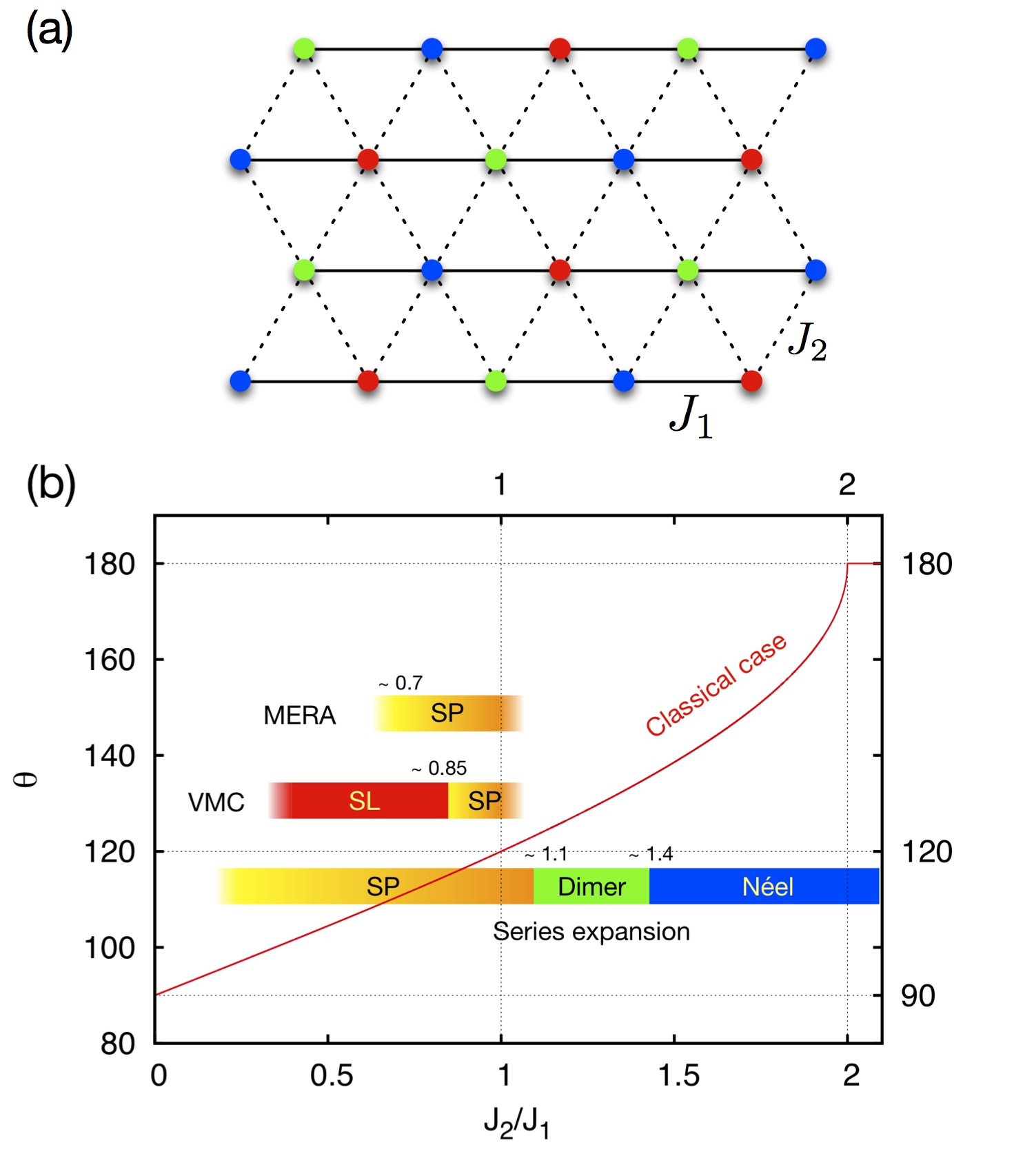}
  \caption{(Color online) (a) Spatially anisotropic triangular
    lattice with two groups of interactions, denoted by
    dotted and solid links. $J_1$ and $J_2$ denote the coupling
    coefficients on each link.  (b) The magnetic
    moment twist angle along the $J_2$ axis of the classical model (solid line) and
    the proposed phase diagrams of the quantum model (horizontal
    strips) on the spatially anisotropic triangular lattice. Here, SP
    and SL denote the spiral and spin-liquid phase, respectively.  }
  \figlabel{model}
\end{figure}
The simplest effective model of spin degrees in Mott insulators is the
$S=1/2$ antiferromagnetic Heisenberg model on a triangular
lattice. Since the triangular layer in a real material is distorted,
we have two groups of Heisenberg interactions\cite{Powell:2011ce}. 
Figure~\ref{fig:model}(a) shows these two kinds of interactions as solid
and dotted links. The Hamiltonian is written as
\begin{equation}
  \eqlabel{Ham}
  \mathcal{H}= J_1\sum_{\langle ij \rangle} \mathbf{S}_i\cdot \mathbf{S}_j
  + J_2\sum_{\langle \langle ij \rangle \rangle} \mathbf{S}_i\cdot \mathbf{S}_j,
\end{equation}
where $\langle ij \rangle$ and $\langle\langle ij \rangle\rangle$
denote pairs of sites on solid and dotted links in \figref{model}(a),
respectively. The coupling coefficients $J_1$ and $J_2$ are
positive. The ratio $J_2/J_1$ in real materials varies widely, from
$\frac13$ to $1$. For example\cite{{Powell:2011ce}}, $J_2/J_1$ for
$\kappa$-(BEDT-TTF)${}_2$ Cu${}_2$ (CN)${}_3$ was estimated to be
close to 1, and that for Cs${}_2$CuCl${}_4$ was estimated to be about
$\frac13$.
\begin{figure}[ht]
  \centering
  \includegraphics[width=0.4\textwidth]{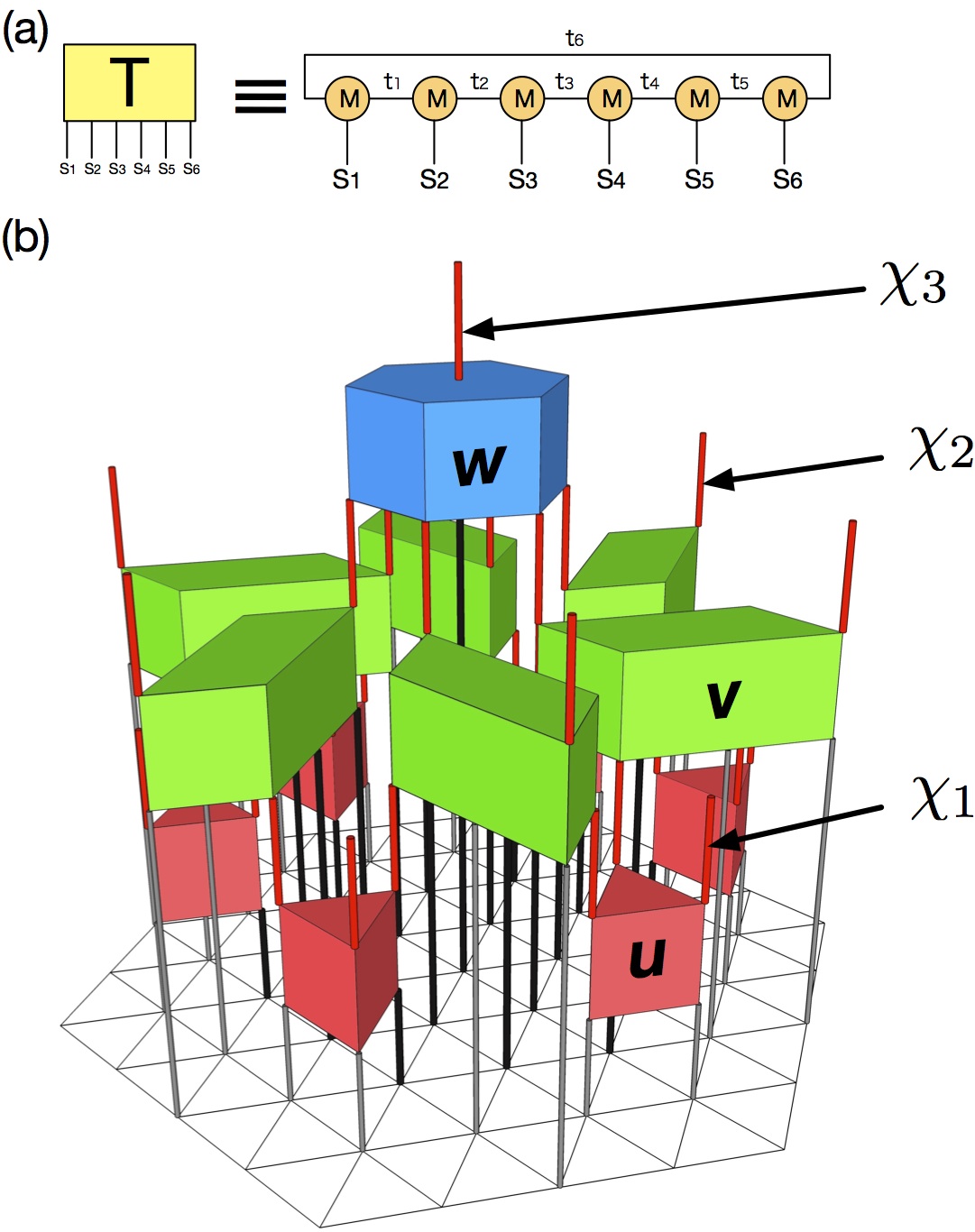}
  \caption{(Color online) (a) Matrix product state for six sites:
    $T^{\mathrm{MPS}}_{s_1,\cdots,s_6}$.  (b) Tensor network with
    entanglement renormalization on a triangular lattice.  Solid (red)
    lines represent tensor contractions for two connected tensor
    indices.}
  \figlabel{model-TN}
\end{figure}

The model of \eqref{Ham} interpolates among independent chains
($J_2=0$), the fully frustrated triangular lattice ($J_1=J_2$), and
the unfrustrated square lattice ($J_1=0$).  Geometrical frustrations
are present throughout the model, except at two special points($J_1=0$
and $J_2=0$).  In the classical case, the ground state can be solved
exactly: There are two long-range order phases at zero temperature
[see \figref{model}(b)]: a N\'eel state on a square lattice for
$J_2/J_1 \ge 2$ and a spiral state with a smoothly changing wave
number for $J_2/J_1 \le 2$. However, the phase diagram in the quantum
model cannot be solved analytically.  Thus, we have used various
numerical or approximate methods. First, the ground state at the
isotropic point ($J_2/J_1=1$) has been studied. For example, both
exact diagonalization \cite{{Bernu:1994iw}} and \myabb{density matrix
  renormalization group}{DMRG} calculations \cite{{White:2007da}}
reveal a $120^\circ$ magnetic ordered ground state, whose wave number
is equal to that in the classical model. Although quantum fluctuations
reduce the magnetic moment magnitude, the $120^\circ$ magnetic ordered
state was confirmed. However, the stability of the magnetic ordered
state in the anisotropic region is controversial. Yunoki and
Sorella\cite{Yunoki:2006hn} reported a disordered state for $J_2/J_1
\lesssim 0.79$ by using a \myabb{variational Monte Carlo}{VMC}
technique. In addition, Heidarian et al.\cite{Heidarian:2009gd}
reported the disappearance of magnetic long-range order at $J_2/J_1
\sim 0.85$ by using another VMC technique. Weng et al.
\cite{Weng:2006ck} also reported a similar disordered state for
$J_2/J_1 \le 0.78$ by using the DMRG method. Reuther and
Thomale\cite{Reuther:2011gv} reported a disordered state with
collinear antiferromagnetic stripe fluctuations for $J_2/J_1 < 0.7
\sim 0.9$ by using the pseudofermion functional renormalization group
method.  Thus, the disordered behavior in real materials may be
captured by these states. However, some reports are contradictory, as
shown in \figref{model}(b).  Zheng et al. proposed a spiral phase for
$0 < J_2/J_1 < 1.11$ by using a series expansion
method\cite{Zheng:1999cu}. Weichselbaum and White
\cite{Weichselbaum:2011hq} also reported a long-rang magnetic
correlation with an incommensurate wave vector in the whole region of
$0 < J_2/J_1 \le 1$ by using the DMRG method with different boundary
conditions. The renormalization group analyses\cite{Starykh:2007gj,
  Ghamari:2011dn} suggest a direct transition from spiral to collinear
antiferromagnetic order at $J_2/J_1 \lesssim 0.3$.  Therefore, the
stability of the spiral state in the quantum case is crucial for
understanding the physical behavior of real materials.

In this study, a new numerical approach was used to calculate the
ground state. Usually, \myabb{quantum Monte Carlo}{QMC} methods are
powerful tools for two-dimensional quantum models, because they are
unbiased. However, the weight of QMC samples can be negative in
frustrated quantum magnets, leading to a cancellation in sign, and the
accuracy of simulations fatally decreases (this is the so-called sign
problem). Exact diagonalization can only be applied to small
systems. Thus, a variational method has been chosen in this study. In
particular, the key point of calculations in this study is the trial
wave function. It is based on a tensor network with
\myabb{entanglement renormalization}{ER}\cite{{Vidal:2007kx}}. A
tensor network is a theoretical tool in the field of quantum
information to describe a quantum state. By modifying the network
structure, we can freely design the structure of entanglements that
mean quantum correlations in a quantum state. In general, the
entanglement entropy of a subsystem is proportional to the area of the
boundary\cite{{Srednicki:1993dl}}. A tensor network with ER also obeys
the area law of entanglement entropy\cite{{Barthel:2010fx}}.  Though
it only has a bias owing to the particular network structure used in
the calculation, systematic error can be controlled, in principle, by
increasing the dimensions of tensor indices. Therefore, the tensor
network method is regarded as one of the most promising techniques for
treating numerically unsolved problems such as the present one.
Unfortunately, successful applications to quantum frustrated magnets
in two dimensions are very few \cite{{Evenbly:2009ec},
  {Evenbly:2010hh}}. In what follows, we demonstrate the usefulness of
ER by applying it to the model of \eqref{Ham} to clarify the nature of
its ground state.

By using the ER tensor networks shown in \figref{model-TN}(b),
the spiral state with incommensurate wave numbers for $0.7
\le J_2/J_1 < 1$ that overlaps with those of the disordered (spin
liquid) phase reported in previous works \cite{Yunoki:2006hn,
  Heidarian:2009gd, Weng:2006ck} was confirmed [see \figref{model}(b)]. In
the numerical results, quantum fluctuations weaken the effective coupling between
chains, but the long-range magnetic order remains in the thermodynamic
limit. In addition, the incommensurate wave number is
in good agreement with that obtained by the series expansion
method\cite{Zheng:1999cu}. Since these two approaches are different,
the results of this study provide strong evidence for the stable spiral phase.

The paper is organized as follows: In Sec. \ref{sec:tensor-network}, a
tensor network with ER designed for triangular lattice models will be
briefly introduced. In Sec. \ref{sec:results}, numerical calculations
of the $S=1/2$ antiferromagnetic Heisenberg model on a spatially
anisotropic triangular lattice will be reported. In
Sec. \ref{sec:conclusions}, the results will be summarized.

\section{Tensor network with ER on a triangular lattice}
\label{sec:tensor-network}
\subsection{Tensor network}
Formally, the probability amplitudes of a wave function $\vert \psi
\rangle$ can be regarded by a rank-$N$ tensor $T$ as $\langle s_1,
\cdots, s_N \vert \psi \rangle \equiv T_{s_1, \cdots, s_N}$, where $N$
denotes the number of sites. However, we cannot treat a large-$N$-site
system by using a tensor, because the number of elements in a rank-$N$
tensor exponentially increases. To avoid the exponential increase, we
replace the original large-rank tensor by using a set of tensor
contractions of small-rank tensors.

The tensor contractions can be drawn as a network.  Thus, this type of
wave function is called a \textit{tensor network wave function} or,
simply, a \textit{tensor network}. The node of the network denotes a
tensor, and the leg of the node denotes a tensor index. An edge
connecting two legs represents a tensor contraction for the two
corresponding indices. For example, \figref{model-TN}(a) shows the
tensor network for the probability amplitude as
$T^{\mathrm{MPS}}_{s_1, \cdots, s_6} \equiv \sum_{t_1, \cdots, t_6}
M_{t_6,s_1,t_1}\cdots M_{t_{5}, s_{6},t_{6}}$.  The one-dimensional
tensor network is called a \myabb{matrix product state}{MPS}. It is
the general form of the wave function used in DMRG calculations.

\subsection{ER on triangular lattices}
Various types of tensor networks have been proposed for many-body
quantum systems. The structure of the network affects entanglements in
a tensor network state. For example, the MPS breaks the area law of
entanglement entropy in more than two dimensions. Thus, in principle,
it is not suitable for capturing the quantum state in two-dimensional
quantum systems. To construct a tensor network for a triangular
lattice model, we use a coarse-graining transformation removing
short-range entanglements between coarse-grained regions. This is the
ER method proposed by Vidal\cite{Vidal:2007kx}. In particular, since
the tensor network with ER obeys the area law of entanglement entropy,
it can describe a quantum state with large entanglements in principle.

\begin{figure}
  \centering
  \includegraphics[width=0.45\textwidth]{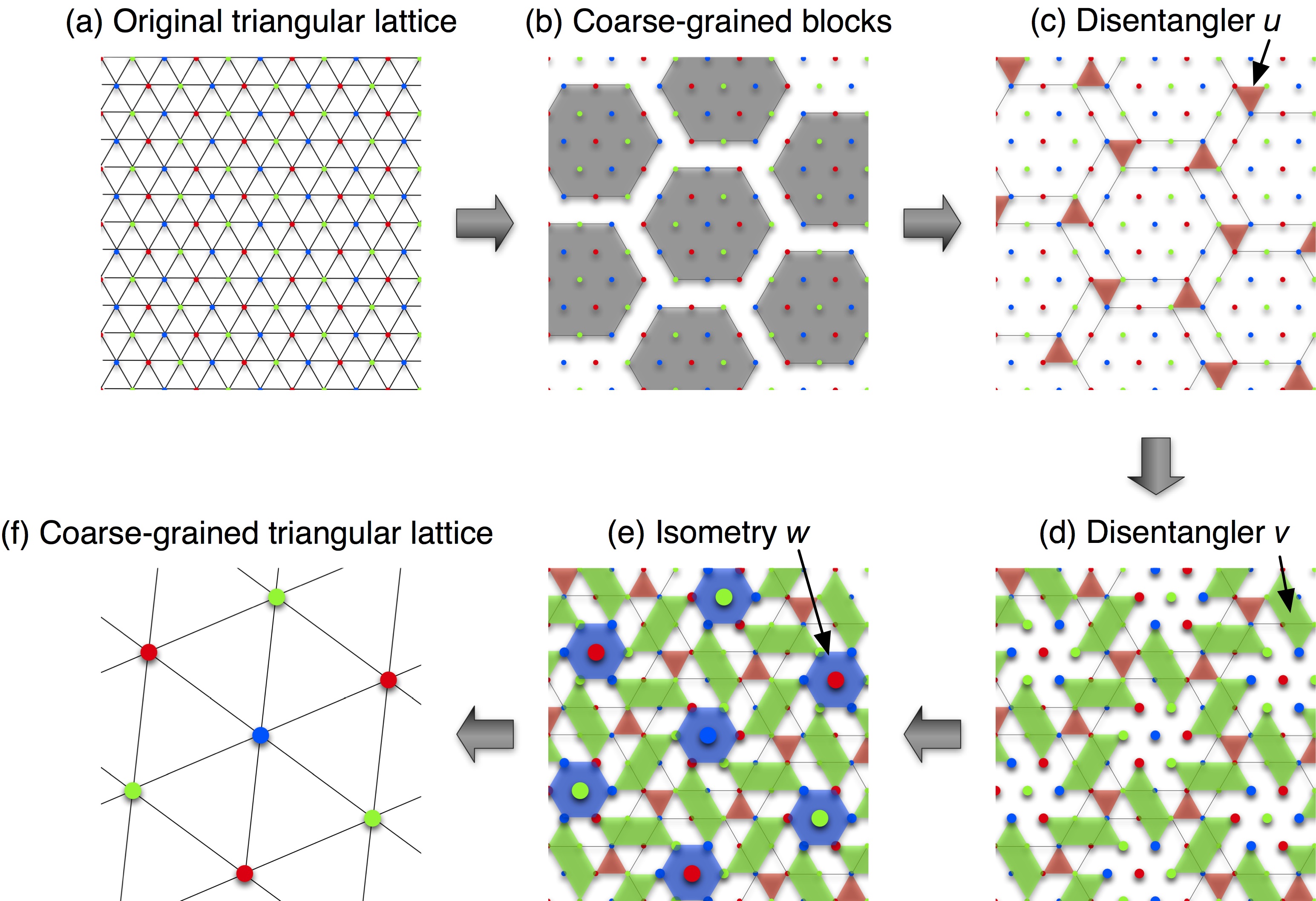}
  \caption{(Color online) Coarse-graining transformation for a
    triangular lattice model.}
  \figlabel{ER}
\end{figure}

No systematic studies have been conducted on the optimal network
structure. The empirical rule is that it should decrease the
entanglement between coarse-grained regions as much as possible and,
at the same time, keep the computational cost of tensor contractions
reasonable. Figure~\ref{fig:model-TN}(b) shows a suitable ER tensor
network for triangular lattice models. It transforms a triangular
lattice [\figref{ER}(a)] to a coarse-grained one
[\figref{ER}(f)]. After the transformation, the number of sites
decreases by a factor of $19$. The coarse-grained unit cell is the
filled gray hexagon in \figref{ER}(b). As shown in
\figref{model-TN}(b), the network consists of three sublayers. Each
sublayer is occupied by a single type of tensor: $u$(red), $v$(green),
and $w$(blue) from the bottom sublayer to the top sublayer,
respectively, as shown in \figref{model-TN}(b) and \figref{ER}.
Tensors $u$ and $v$ are called \textit{disentanglers}, because their
purpose is to decrease short-range entanglements between
coarse-grained regions. The tensors have upper and lower legs. An
upper leg in one sublayer is connected to the lower leg of a tensor in
the higher sublayer.  Disentangler $u$ has three lower legs and three
upper ones. Disentangler $v$ has six lower legs and two upper ones.
Tensor $w$ transforms seven sites to one site; this process is called
\textit{isometry}. In principle, the ER can be applied iteratively,
and we usually finish the ERs corresponding to the top tensor on the
last coarse-grained lattice, which is a simple isometry. In
particular, the tensor network with multiple-level ERs is called a
\myabb{multiscale entanglement renormalization ansatz}{MERA}.

\subsection{Computational costs of MERA}
All tensors in MERA are isometric: $\sum_k (T^i_k)^* T^j_k = \delta_{ij}$,
where $T^i_j$ denotes the tensor's element with index $i$ ($j$) of
upper (lower) legs. Because of the isometric property, an expectation
value of a local operator can be evaluated on a subnetwork that is
finite and much smaller than the whole network, in most
applications. This subnetwork is called a \textit{causal
  cone}\cite{Vidal:2007kx, Evenbly:2009bk, Giovannetti:2009ku}.
Figure~\ref{fig:CC} shows a causal cone for the expectation value of
an operator on a triangle plaquette of nearest neighbor sites.  The
number of tensors in a causal cone is proportional only to the
logarithm of system size. Thus, the computational cost depends on the
system size only weakly, compared to the exponential growth that is
naturally expected. It only increases by a polynomial of dimensions of
tensor indices.
\begin{figure}[ht]
  \centering
  \includegraphics[width=0.5\textwidth]{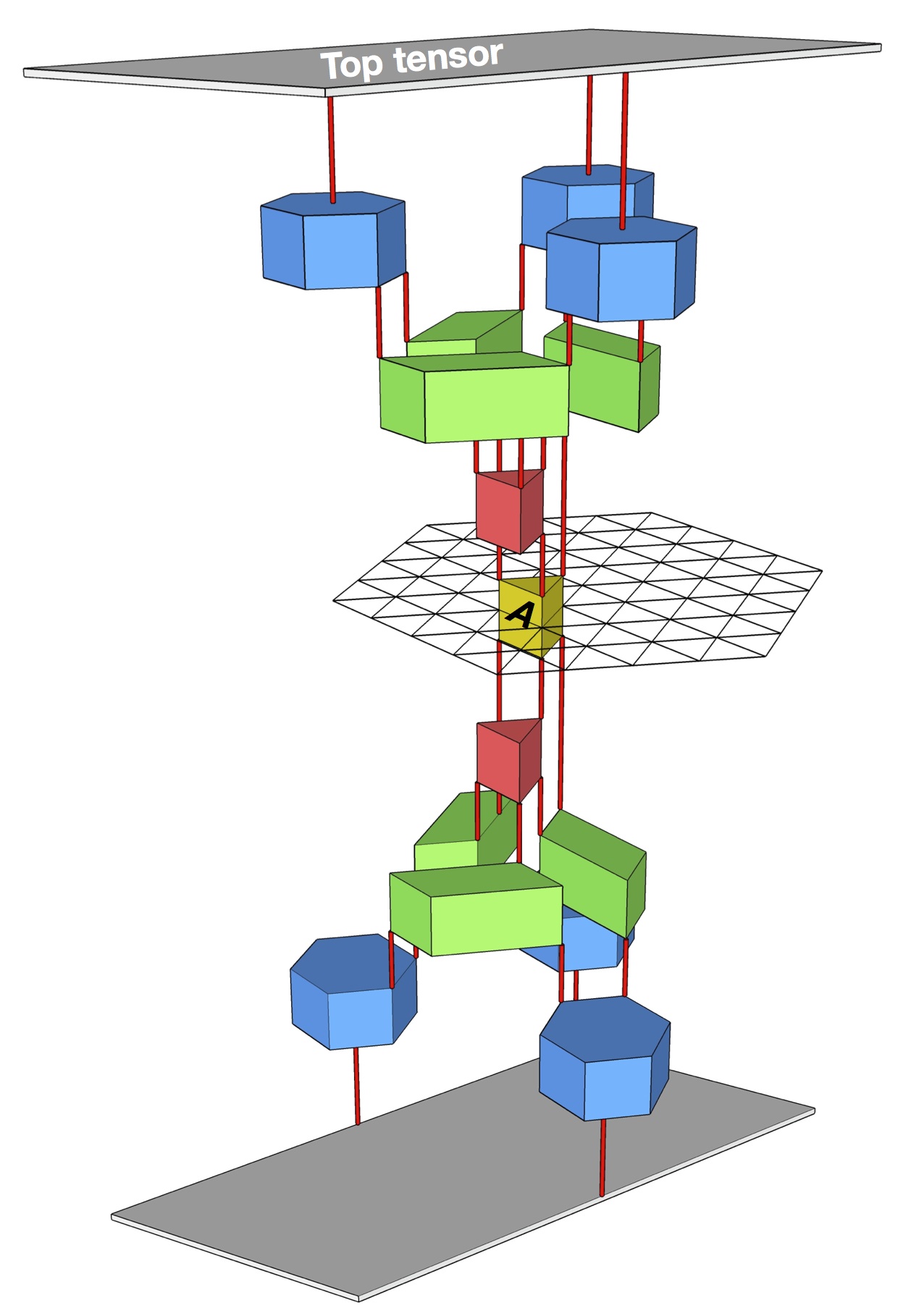}
  \caption{(Color online) Causal cone for an expectation value of a
    local operator $A$ on a triangle plaquette of nearest neighbor
    sites $i$, $j$, and $k$: $\langle \psi \vert A(i,j,k) \vert \psi
    \rangle$. Here $\vert\psi\rangle$ is represented by the tensor
    network with an ER level defined in \figref{model-TN}(b) and a top
    tensor that covers six sites as in \figref{L114}. The local
    operator $A$ is drawn as a (yellow) tensor that has three upper and three
    lower legs.  The upper and lower parts of the (yellow) tensor
    are subnetworks from $\vert \psi \rangle$ and $\langle \psi
    \vert$, respectively.  We only draw tensor contractions between
    unpaired tensors as solid (red) lines. A tensor's index without a
    solid (red) line in the upper part is always connected to that at
    the same position in the lower part. The width of a causal cone is
    defined by the number of solid (red) lines at the same height. The
    maximum width of this causal cone is six.  }
  \figlabel{CC}
\end{figure}

We assume that tensor legs at the same ``height'' have the same
dimensions. Then the size of tensors in MERA can be specified by only
an integer set as $(\chi_1, \chi_2, \chi_3)$ in \figref{model-TN}(b),
where $\chi_i$ is the dimension of the upper index of tensors in the
$i$th sublayer. The computational cost of the expectation value of
operators on a triangle plaquette becomes a polynomial of these
integers. As shown in \figref{CC}, all tensors except the local
operator drawn as (yellow) tensor $A$ are paired in a causal cone.  We
first calculate tensor contractions between paired tensors. Then, the
causal cone is transformed to a tensor network that has a half height.
Each tensor in the new tensor network is defined by paired tensors in
the original network.  If and only if an edge connects unpaired
tensors, the edge remains in the new tensor network, as shown by the
solid (red) lines in \figref{CC}.  Thus, the shape is similar to the
upper part of the original one.  The number of remaining lines at the
same height in the new tensor network is called the width of the
causal cone. Usually, we calculate this tensor network from the local
operator drawn as (yellow) tensor $A$ in \figref{CC}. The maximum
number of indices of intermediate tensors is roughly double the
maximum width of the causal cone. Thus, the memory size needed for
calculating a causal cone rapidly increases with increasing dimensions
of tensor indices.  In fact, the maximum width of the causal cone in
\figref{CC} is six. Since we use multithreaded subroutines for tensor
contractions, the total memory size needed to calculate a causal cone
is limited by the memory size of a computational node, which strictly
limits the maximum dimensions of indices. In addition, the maximum
polynomial degree for the computation of the causal cone is also
larger than the double maximum width of causal cone. For example, the
maximum polynomial degree in \figref{CC} is $14$. Since the
Hamiltonian of \eqref{Ham} is written as the summation of local
Hamiltonians on triangle plaquettes of nearest neighbor sites, the
main part of the variational method can be decomposed into
calculations of independent causal cones corresponding to local
Hamiltonians. Thus, this part can be perfectly parallelized. Main
calculations have been done using the facilities of the Supercomputer
Center, Institute for Solid State Physics, University of Tokyo.  In
the largest case for the tensor network with two ER levels, 256 nodes
were used.

\section{Numerical results obtained using a tensor network with ER}
\label{sec:results}

\subsection{Isotropic triangular lattice}
\label{ssec:ground-state-at-isotropic}
First, we calculate ground states of finite and infinite systems for
$J_1=J_2$. The $120^\circ$ magnetic ordered state at the isotropic
point has been confirmed by previous works (see Table III in
Ref.~\onlinecite{Zheng:2006iy}). The purpose of this calculation is to
test the variational wave function defined in the previous section and
to see the behavior for an S=1/2 antiferromagnetic Heisenberg model on
a triangular lattice.

\subsubsection{Tensor network}
The wave function consists of the single ER level in \figref{ER} with
a top tensor. The top tensor covers six coarse-grained sites after the
ER. Thus, this tensor network structure is applied to $N=6 \times
19=114$ sites. Figure~\ref{fig:L114} shows the tensor network
structure.  Large solid circles denote the positions of coarse-grained
sites.  We put the top tensor on the parallelogram frame in
\figref{L114}. We apply this tensor network structure to both finite
and infinite lattices. In this paper, we call the former the
\myabb{periodic boundary condition}{PBC} scheme and the latter the
infinite-size scheme, respectively.

In the PBC scheme, the total number of sites is just $6 \times 19 =
114$. We set a skew PBC so that all parallelogram frames in
\figref{L114} are the same. We notice that this PBC is consistent with
a three-sublattice structure of a triangular lattice. In contrast, in
the infinite-size scheme, we arrange the same tensor network
structure, with $114$ sites per unit cell on an infinite lattice.
Then the top layer is defined as the product state by top tensors. In
addition, we assume that the tensors at the corresponding positions in
all repeated units are the same. Thus, we can define a wave function
by the finite set of tensors for the infinite lattice. This type of
MERA is called a finite-correlation MERA\cite{Evenbly:2009bk}, because
reduced correlations become exactly zero for large distances. The
distance limit for finite reduced correlations is roughly the size of
the unit cell. The main difference between the two schemes lies in the
causal cones. In the PBC scheme, all the causal cones are limited to
one unit cell, because of the PBC.  However, in the infinite-size
scheme, some causal cones extend into multiple unit cells. Thus, the
computational time for the infinite-size scheme may be longer than
that for the PBC scheme.
\begin{figure}
  \centering
  \includegraphics[width=0.45\textwidth]{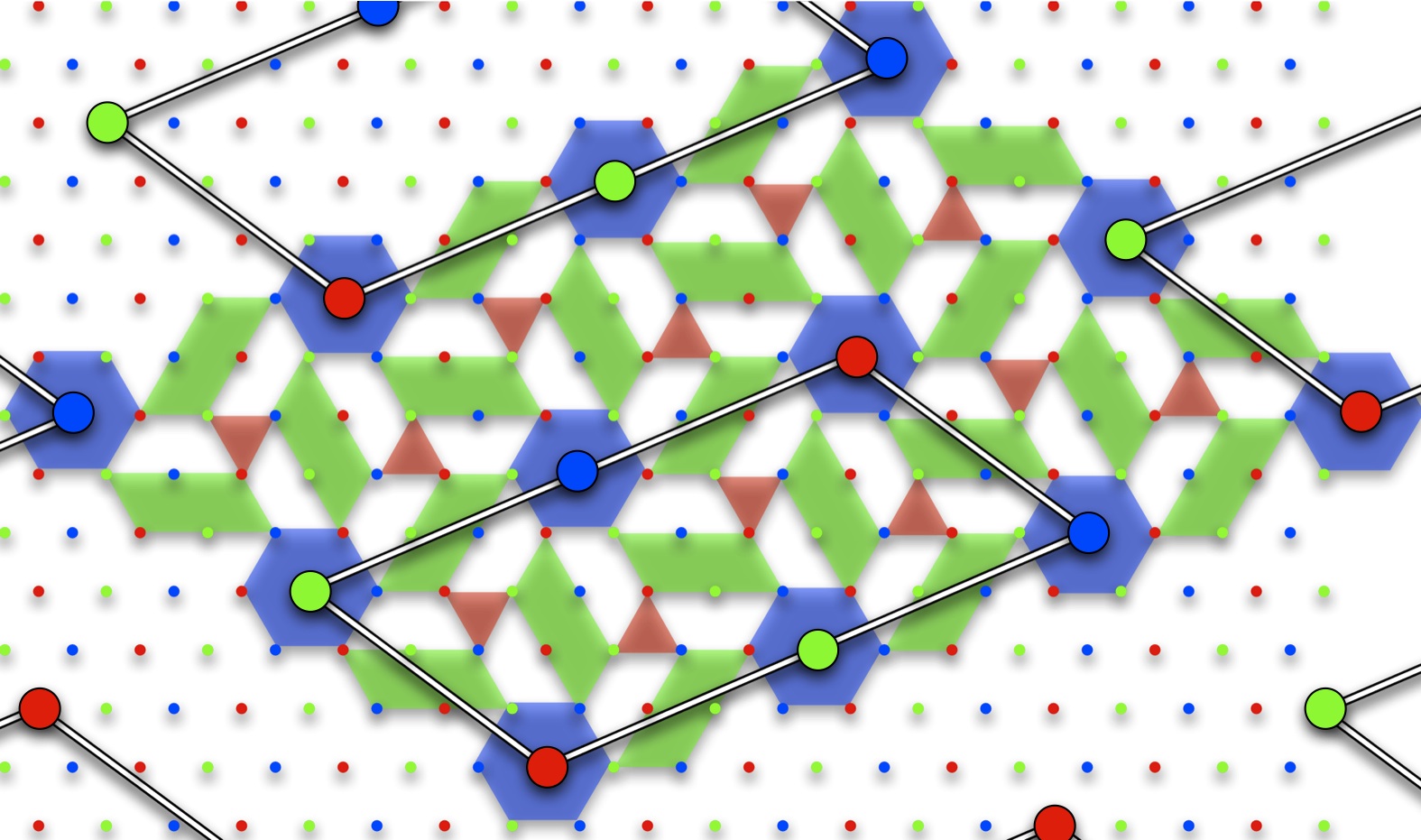}
  \caption{(Color online) Tensor network structure with a single ER
    level and a top tensor for six coarse-grained sites. Big solid
    circles denote positions of sites on the coarse-grained
    lattice. The top tensor is put on the parallelogram frame.  All
    parallelogram frames are the same by a skew periodic arrangement.}
\figlabel{L114}
\end{figure}

We assume that all tensors are independent within the unit cell to
have more variational freedom with less bias.  They were optimized to
minimize the total energy of the tensor network states. The tensors
are iteratively updated by the singular value decomposition method
\cite{Evenbly:2009bk}. Although this minimization problem may have
some local minimum states, stable results starting from
random initial tensors were obtained.

\subsubsection{Energy}
The ER tensor network for this study has the spatial structure shown
in \figref{ER}. As we mentioned above, if we use the finite dimension
of the tensor index, the network-structure bias may cause a
significant systematic error.  To check whether or not this is the
case, we try several sets of dimensions of tensor indices.
Figure~\ref{fig:ene_dist_100100} shows local energies on finite and
infinite lattices for various tensor sizes. The value of the local
energy on a triangle plaquette defined by nearest neighbor sites $i$,
$j$, and $k$, $\mathbf{S}_i \cdot \mathbf{S}_j + \mathbf{S}_j \cdot
\mathbf{S}_k + \mathbf{S}_k \cdot \mathbf{S}_i$, is shown on a color
scale. In the $120^\circ$ state, which we believe is the ground state
in the present case, the local energy is homogeneous. Therefore, we
expect that the whole system should be uniformly colored if the error
of the calculation is sufficiently small.
Figures~\ref{fig:ene_dist_100100}(a) and \ref{fig:ene_dist_100100}(c)
show the results of small tensor size, $(\chi_1, \chi_2,
\chi_3)=(2,2,2)$, for PBC and infinite-size schemes of $N=114$,
respectively. There are clear spatial patterns resulting from the
structure of ER in both cases. Increasing the tensor size should
improve the quality of the tensor network states. In fact, as shown in
Figs. \ref{fig:ene_dist_100100}(b) and \ref{fig:ene_dist_100100}(d),
the patterns are clearly more homogeneous than for small tensors. We
notice that the patterns in the infinite-size scheme are quite similar
to those of the PBC scheme. Thus the assumption of direct product
states for infinite systems does not affect the spatial pattern of
local energies in the tensor network states.
\begin{figure}
  \centering
  \includegraphics[width=0.45\textwidth]{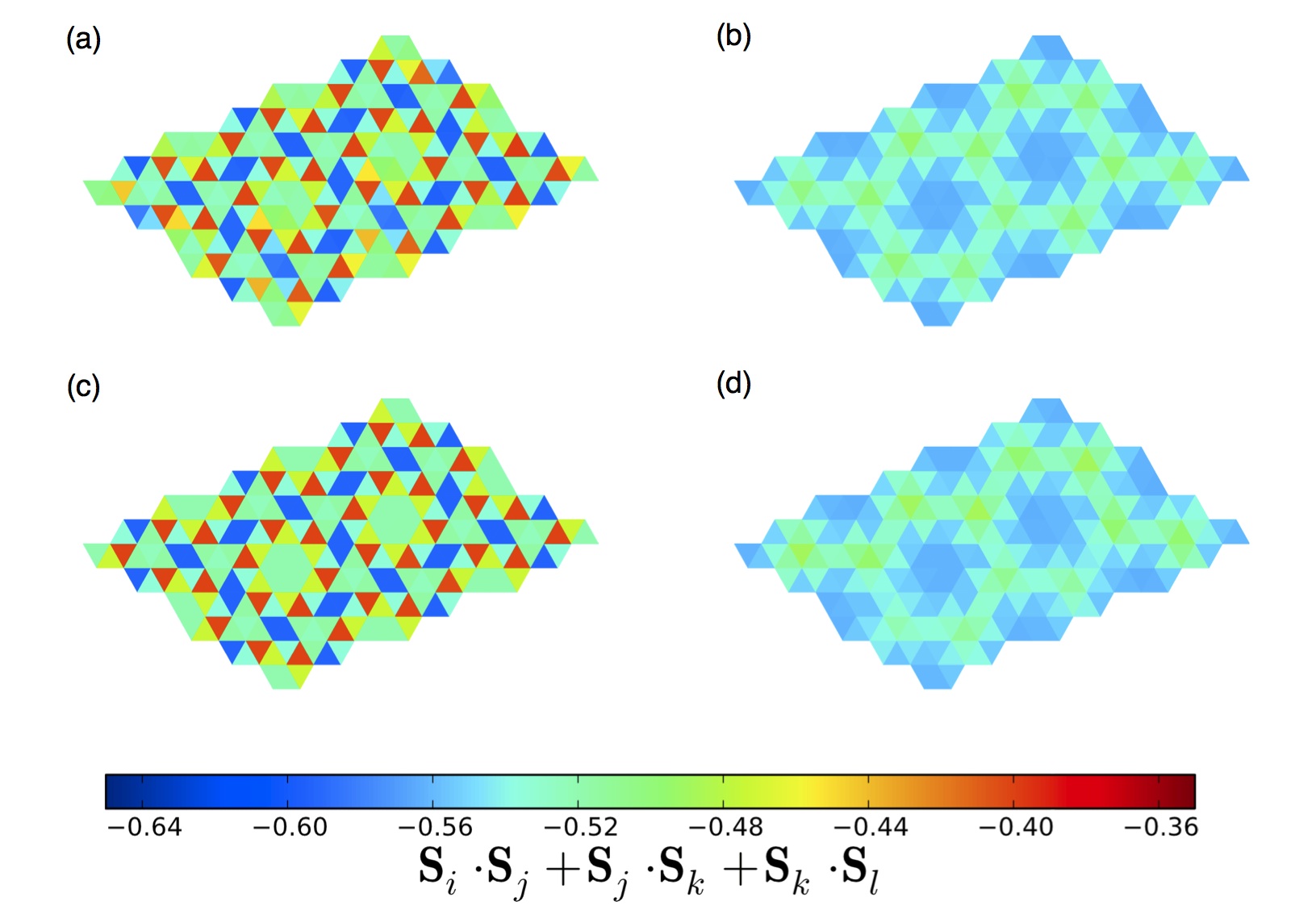}
  \caption{(color online) Local energies on the triangular
    plaquettes. The value of the local energy on a triangle plaquette
    defined by nearest neighbor sites $i$, $j$ and $k$, $\mathbf{S}_i
    \cdot \mathbf{S}_j + \mathbf{S}_j \cdot \mathbf{S}_k +
    \mathbf{S}_k \cdot \mathbf{S}_i$, is shown by the color scale. The
    bottom bar shows the correspondence between the value of local
    energy and color.  Results of the PBC scheme are shown in (a) and
    (b). Those of the infinite-size scheme are shown in (c) and (d). The
    size of tensors in (a) and (c) is $(\chi_1, \chi_2,
    \chi_3)=(2,2,2)$.  In (b) and (d), $(\chi_1, \chi_2,
    \chi_3)=(2,8,8)$.  }
  \figlabel{ene_dist_100100}
\end{figure}

Figure~\ref{fig:ene_ent_100100} shows the energy per site, $E$, for
the tensor sets $(\chi_1, \chi_2, \chi_3)=(2,2,2),\ (2,4,4),\
(2,8,4),$ and $(2,8,8)$ for both PBC and infinite-size schemes. When
the tensor size increases, the energy of tensor networks is indeed
improved.  The lowest energy per site at $(2,8,8)$ is $E_{\rm PBC}
(N=114)=-0.54181$ and $E_{\rm inf} (N=114)=-0.54086$.  These values
compare well with results obtained from other methods (see Table III
in Ref.~\onlinecite{Zheng:2006iy}). In particular, the result of a
\myabb{Green's function quantum Monte Carlo}{GFQMC} calculation with
\myabb{stochastic reconfiguration}{SR}\cite{Capriotti:1999kk} is
$E(N=144) = -0.5472(2)$. Direct comparison may be difficult, because
the skew boundary in this study is not equal to the periodic one used by the authors
of Ref.~\onlinecite{Capriotti:1999kk} and the lattice ($N=114$) in this study is
smaller than their lattice ($N=144$). However, the result of this study is close to
their result (:$1\%$ lower than its). The result for the
infinite-size scheme also agrees well with previous estimates for the
thermodynamic limit. In particular, it compares favorably to that of a
series expansion\cite{Zheng:2006iy}, $E = -0.5502(4)$, and to that
from a GFQMC with SR calculation\cite{Capriotti:1999kk}, $-0.5458(1)$,
in the thermodynamic limit.
\begin{figure}
  \centering
  \includegraphics[width=0.45\textwidth]{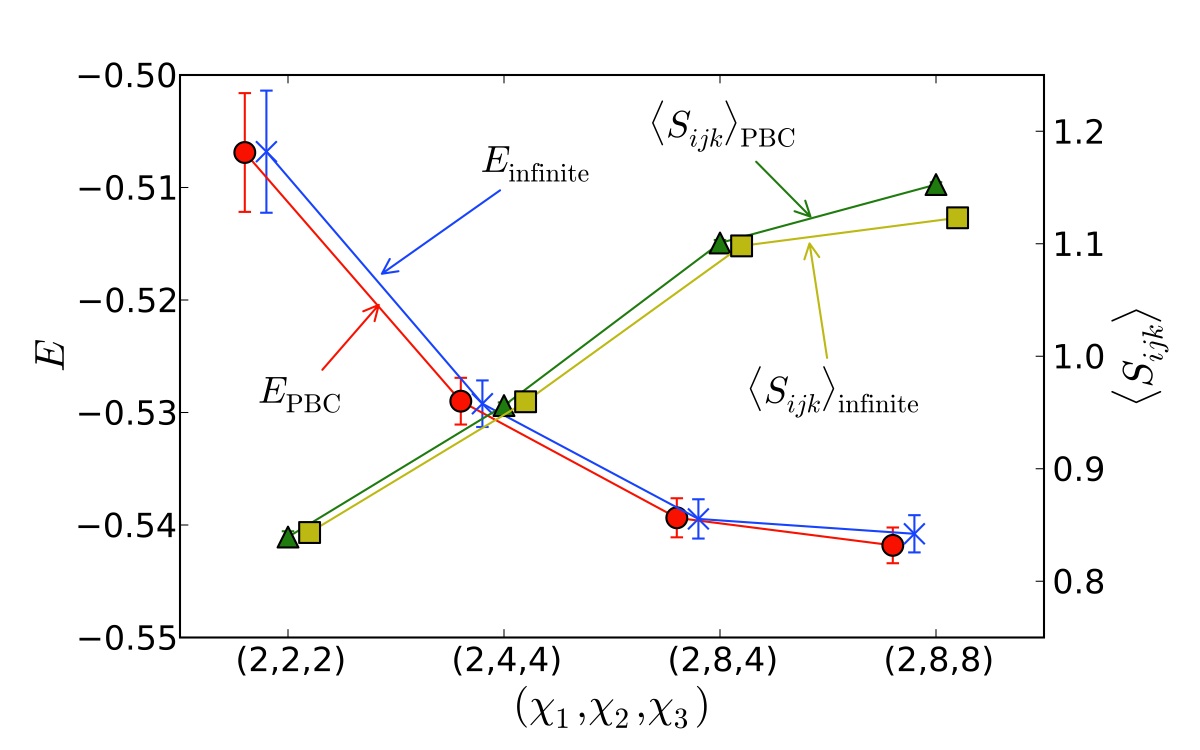}
  \caption{(Color online) Energy per site, $E$, and the average of
    entanglement entropies on triangle plaquettes, $\langle
    S_{ijk}\rangle$, for PBC and infinite-size schemes
    ($N=114$). Circles (red) and crosses (blue) denote $E$ for PBC and
    infinite-size schemes, respectively.  Triangles (green) and
    squares (yellow) denote $\langle S_{ijk} \rangle$ for PBC and
    infinite-size schemes, respectively. The error bar of $E$ shows
    the deviation of local energies on triangular plaquettes.  }
  \figlabel{ene_ent_100100}
\end{figure}

\subsubsection{Entanglement entropy}
\begin{figure}
  \centering
  \includegraphics[width=0.45\textwidth]{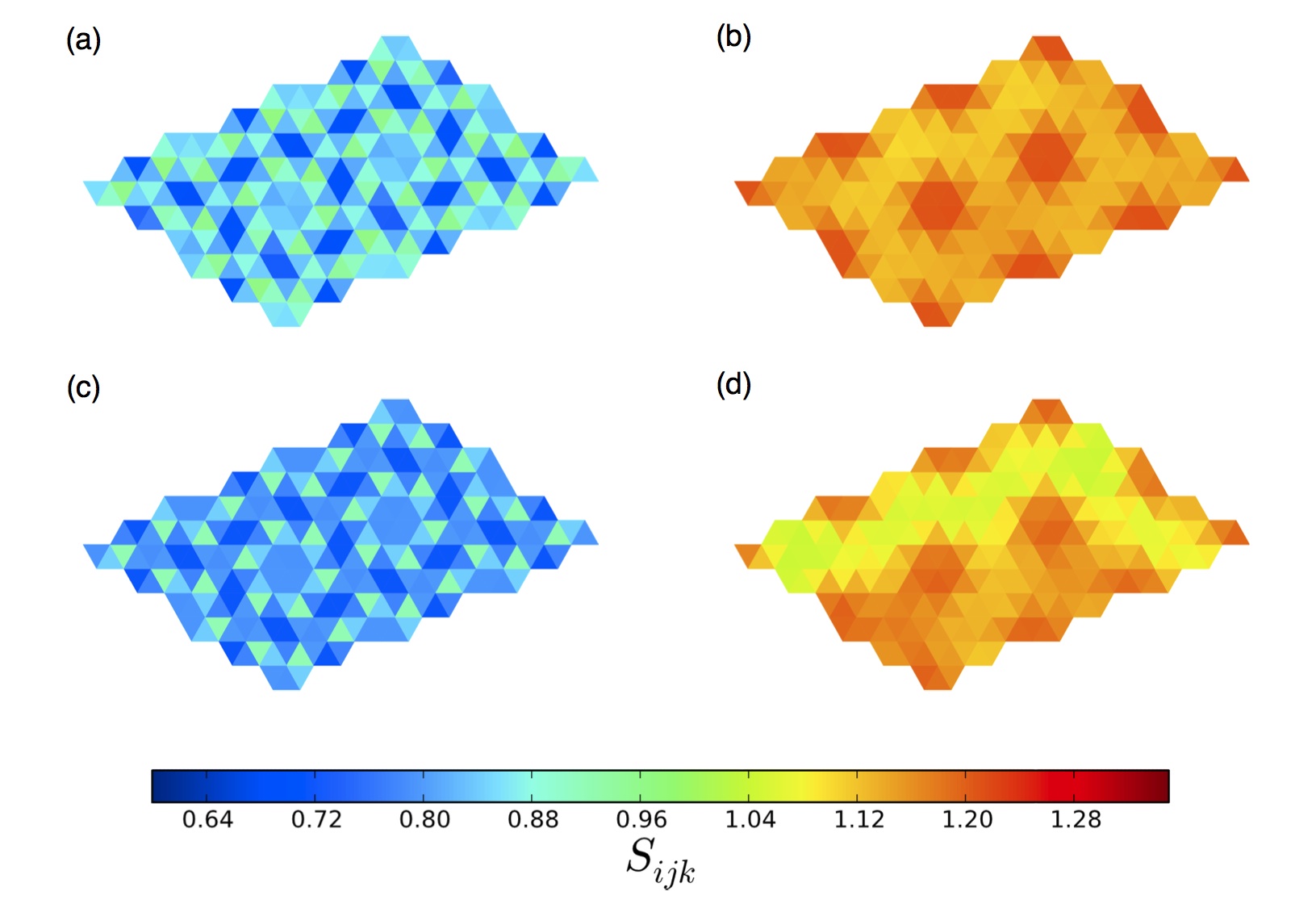}
  \caption{(color online) Entanglement entropy of triangular
    plaquettes defined by nearest neighbor sites $i$, $j$, and $k$ in
    ER tensor network states. The bottom bar shows the correspondence
    between the value of entanglement entropy $S_{ijk}$ and color. The
    schemes and tensor sizes from (a) to (d) are equal to those from
    (a) to (d) in \figref{ene_dist_100100}, respectively.}
  \figlabel{ent_100100}
\end{figure}
Entanglement entropy measures the quantum correlation between a
considered region and another region. Figure~\ref{fig:ent_100100} shows the
entanglement entropy of a triangular plaquette on a color scale. It is
defined as
\begin{equation}
  \label{eq:entropy}
S_{ijk} \equiv -{\rm Tr} [ \rho_{ijk}\ln \rho_{ijk}],
\end{equation}
where $\rho_{ijk}$ is the reduced density matrix of sites $i$, $j$,
and $k$ on a triangle plaquette. There is clear spatial inhomogeneity
owing to the network-structure bias, as is also seen in the local
energy. The entanglement entropies on the boundary of coarse-grained
regions are lower. When the tensor size increases, the average value
of entanglement entropies also increases, as shown in
\figref{ene_ent_100100}. However, as shown in
Figs.~\ref{fig:ent_100100}(b) and \ref{fig:ent_100100}(d), the spatial
patterns of entanglement entropies are different.  In the
infinite-size scheme, the boundary of unit cell has weak entanglement
entropies. The reason for this is that the top layer is a direct
product state. Although the assumption of a direct product state does
not affect the spatial pattern of local energies, that of entanglement
entropy is more sensitive. Thus, the entanglement entropy may be
useful for checking wave function quality in other cases.

\subsubsection{Magnetization}
\begin{figure}
  \centering
  \includegraphics[width=0.45\textwidth]{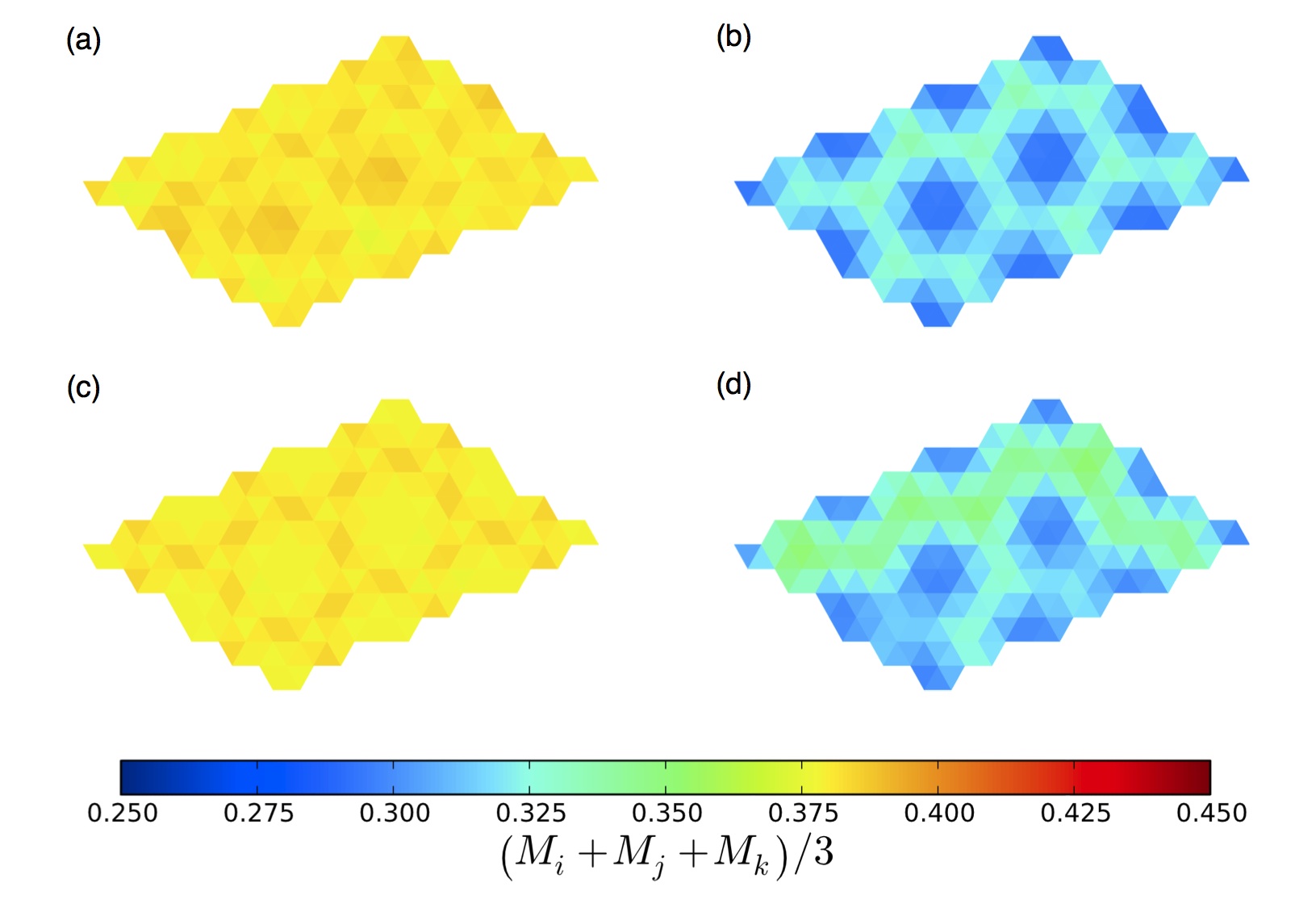}
  \caption{(color online) Averages on-site magnetizations on
    triangular plaquettes. The bottom bar shows the correspondence
    between the value of average on-site magnetizations and color.
    The schemes and tensor sizes from (a) to (d) are equal to those
    from (a) to (d) in \figref{ene_dist_100100}, respectively.}
  \figlabel{mag_100100}
\end{figure}
\begin{figure}
  \centering
  \includegraphics[width=0.45\textwidth]{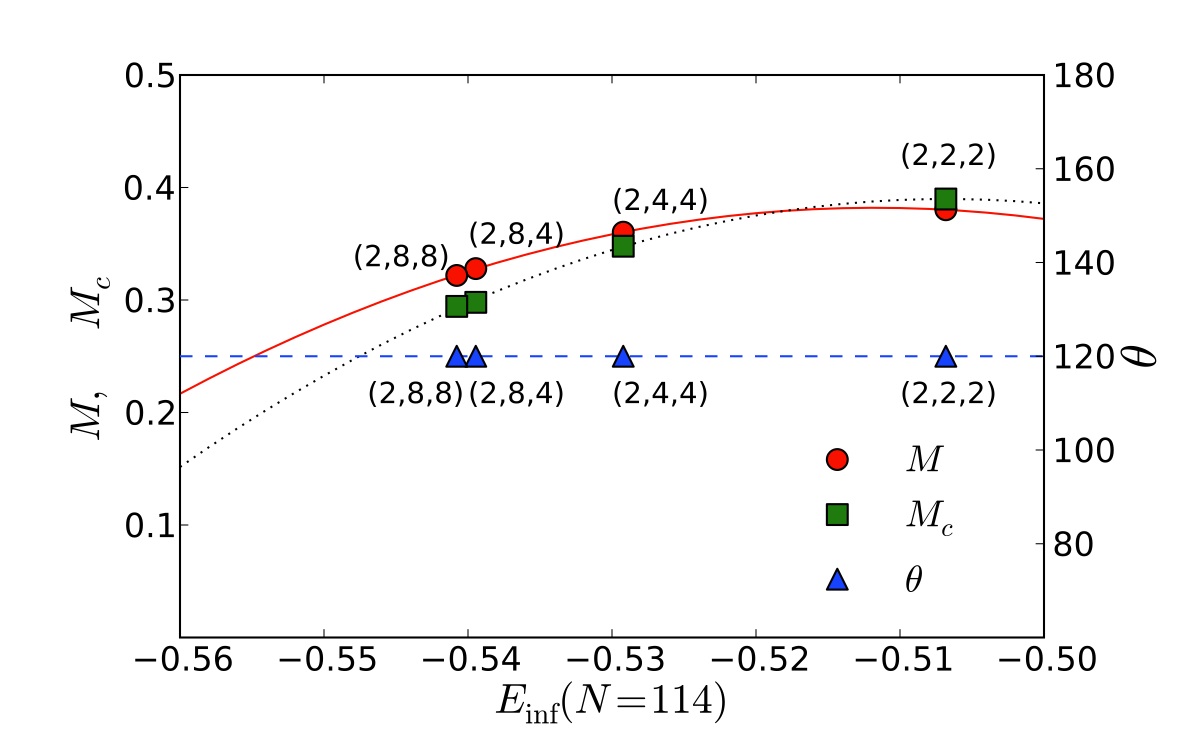}
  \caption{(Color online) Average on-site magnetizations and average
    angles between magnetic moments of nearest neighbor sites.  These
    results are obtained by the infinite-size scheme ($N=114$).
    The horizontal axis denotes energy per site. The left and right vertical
    axes denote the average on-site magnetizations on all sites and
    at the centers of ER, $M$ and $M_c$, and the average angle between
    magnetic moments of nearest neighbor sites, $\theta$, respectively. Circles (red),
    squares (green), and triangles (blue) denote points $(E_{\rm inf}
    (N=114), M)$, $(E_{\rm inf} (N=114), M_c)$, and $(E_{\rm inf}
    (N=114), \theta)$, respectively. Solid (red), dotted (green), and
    dashed (blue) lines are fitting curves for them,
    respectively. The triplet $(\chi_1, \chi_2, \chi_3)$ denotes the tensor size.}
  \figlabel{ene_mag_1010}
\end{figure}

The magnetization on a site $i$ is defined as 
\begin{equation}
  \label{eq:on-site-magnetization}
M_i\equiv\sqrt{\langle \mathbf{S}_i \rangle \cdot
    \langle \mathbf{S}_i\rangle},
\end{equation}
where $\langle \cdot \rangle$ denotes the expectation value of the
operator by a variational wave function.  Figure~\ref{fig:mag_100100}
shows the average on-site magnetizations on a triangular
plaquette. They depend on the tensor size, and they show spatial
patterns that depend on the structure of the ER tensor network. In
addition, as seen in the case of entanglement entropy, the
direct-product nature on the top layer also gives rise to spatial
inhomogeneity.  To make the extrapolation possible, we plot estimates
of the magnetization as a function of the corresponding estimates of
energy.  Figure~\ref{fig:ene_mag_1010} shows the average on-site
magnetizations obtained by using the infinite-size scheme ($N=114$),
$M$, for various tensor sizes. The vertical and horizontal axes denote
$M$ and $E_{\rm inf} (N=114)$, respectively. The dependence of $M$ on
the tensor size is high. While the magnetization for the largest
tensor size is $0.322(2)$ [see the left-most (red) circle], we cannot
simply extrapolate it. The solid (red) curve is fitted to points of
$M$. If we trust the $M-E$ curve, and if we take an estimate of
the energy $E \in (0.54, 0.55)$ upon which various previous works
agree, we can conclude that $M \in (0.275, 0.327)$.  However, the
result of this study is clearly larger than previous estimates: $M=0.205(15)$ from
DMRG calculations\cite{White:2007da}, $M=0.205(10)$ by GFQMC with SR
calculations\cite{Capriotti:1999kk}, and $M=0.19(2)$ by series
expansion\cite{Zheng:2006iy}. As shown in \figref{mag_100100}, the
reason for this discrepancy may be the spatial inhomogeneity caused by
disentangling with small tensors.  In DMRG calculations, to suppress
the effect of the boundary condition as pinning the field on boundary
sites, only on-site magnetization at the center of the system was
used\cite{White:2007da}. To suppress the effect of incomplete
disentangling, we also use on-site magnetization only at centers of
ER, $M_c$. Figure~\ref{fig:ene_mag_1010} plots the average on-site
magnetizations at six centers of ER. If we assume the same condition
for extrapolating $M$, we can conclude that $M_c \in (0.232,
0.298)$. This value is significantly close to other estimations.  It
is probable that the estimate of the magnetization in this study is an
overestimate owing to the intrinsic bias of the tensor network that
generally favors states with less entanglement entropy. From this view
point, estimating the magnetization at a position with larger ER may
be more appropriate than simply taking the spatial average.

However, the angle between magnetic moments of nearest neighbor sites
converges, even when the tensor size is small. Figure~\ref{fig:ene_mag_1010}
shows the average angle between magnetic
moments of nearest neighbor sites, $\theta$, for various tensor sizes.
The angle between magnetic moments of sites $i$ and $j$ is defined as
\begin{equation}
  \label{eq:angle}
\theta_{ij} \equiv
\left(\frac{180^\circ}{\pi}\right) \arccos \left[\frac{\langle
    \mathbf{S}_i\rangle \cdot \langle \mathbf{S}_j \rangle}{M_iM_j} \right].  
\end{equation}
All values in \figref{ene_mag_1010} are $120.0(4)$. Therefore, the
ground state of an S=1/2 antiferromagnetic Heisenberg model on an
isotropic triangular lattice is a magnetic ordered state with
$120^\circ$ structure. This result is consistent with results from many previous
works (see Table III in Ref.~\onlinecite{Zheng:2006iy}).

\subsection{Spatially anisotropic triangular lattice of  $J_2 \le J_1$}
\label{ssec:ground-state-of-anisotropic}

Results of variational calculations for a spatially
anisotropic triangular lattice will be reported. We only consider the case of $J_2 \le
J_1$ in this study. Our main interest lies in the robustness of the
spiral magnetic ordered state.

\subsubsection{Tensor network}
As in the classical model, since the wave vector of a spiral magnetic
ordered state may be incommensurate, we have to be careful about the
periodicity in the variational wave function. Because of the
finiteness of the unit cell in the tensor network, the wave vector of
the magnetic ordered state is restricted. This restriction may cause a
strong bias in variational calculations, in contrast to the case of an
isotropic triangular lattice, where the unit cell of the ER tensor
network in this study is perfectly consistent with the three-sublattice ordered
state.

To weaken the finite-size effect of the unit cell, the number of ER
levels was increased. In the MERA tensor network, the size of the unit
cell increases exponentially by the number of ER levels. A tensor
network with two ER levels has been used. The unit cell covers $6
\times 19 \times 19 = 2166$ sites. In addition, we make all tensors in
the unit cell independent. Thus, the wave vector restriction is
relaxed compared to the tensor network with a single ER level. In
detail, the reciprocal vector is written as
\begin{equation}
  \label{eq:r_vec_2}
  \mathbf{k} =
  \frac{l}{4332}
  \left(
    \begin{array}{c}
    33\\
    63\sqrt{3}
  \end{array}
  \right)
  + \frac{m}{4332}
  \left(
    \begin{array}{c}
      41\\
    11\sqrt{3}
  \end{array}
  \right),
\end{equation}
where the number of independent sets, $(l, m)$, is 2166, because of
the skew periodic arrangement of the unit cell in \figref{L114}. This
number is $19$ times that for the single ER case.  Although the number
of independent causal cones also becomes $19$ times greater than
before, parallel computing was used to calculate them.  However,
because of the memory size limit in a computational node, 
the calculations are limited to a tensor of size $\vec{\chi} = (\chi_1,
\chi_2, \chi_3, \chi_4, \chi_5, \chi_6 ) = (2,8,4,4,8,4)$.

\subsubsection{Energy and quantum mutual information}
The variational calculations by PBC and infinite-size
schemes with two ER levels ($N=2166$) have been performed.

As we see in the case of the single ER level, there is spatial
inhomogeneity resulting from the structure bias of ER in both cases.
By increasing the tensor size, we can systematically improve the
quality of the tensor network states as before. In the single-ER
calculation, as shown in \figref{ent_100100}(d), a weak entanglement
region between unit cells exists in the infinite-size scheme. However,
it disappears in the infinite-size scheme with two ER levels. Because
of the large unit cell obtained by the two ER levels, the finite-size
effect of the unit cell for entanglement entropy is sufficiently
removed. In the following, the results of an
infinite lattice will be mainly reported, which directly corresponds to the thermodynamic
limit.

Figure~\ref{fig:ene_two_step_ER} shows the energy per site for
anisotropic cases from $J_2/J_1 = 0.5$ to $1.0$. The tensor sizes in
the tensor network with two ER levels are $\vec{\chi} =
(2,2,2,2,2,2)$, $(2,4,4,4,4,4)$, and $(2,8,4,4,8,4)$.  The value of
energy is improved by increasing the tensor size. In particular, the
results for the MERA tensor network are better than those of VMC
calculations\cite{Yunoki:2006hn} in the region of $J_2/J_1 \ge
0.75$. Even at $J_2/J_1 = 0.7$, the result of MERA by using the
infinite-size scheme with two ER levels ($N=2166$) is a little (0.5\%)
higher than that of VMC calculations\cite{Yunoki:2006hn}. The difference may
removed by the initial condition or by a tensor optimization process.
However, the difference gets worse in the stronger anisotropic
region for $J_2/J_1 < 0.7$. Figure~\ref{fig:ent_aniso} shows the \myabb{quantum
  mutual information}{QMI} of nearest neighbor sites along the $J_1$ and
$J_2$ axes.  QMI represents the quantum correlation of two sites. If
there is only a classical correlation between two sites, QMI is
zero. The QMI of two sites $i$ and $j$ is defined as
\begin{equation}
  \label{eq:qmi}
  I_{ij} \equiv S_i + S_j - S_{ij},
\end{equation}
where $S_i$ and $S_j$ are the entanglement entropy of site $i$ and
$j$, respectively, and $S_{ij}$ is the entanglement entropy of two
sites $i$ and $j$. As shown in \figref{ent_aniso}, when the spatial
anisotropy increases ($J_2/J_1$ decreases), the QMI along the $J_1$
and $J_2$ axes increases and decreases, respectively. However, we
assume an isotropic entanglement structure in the tensor network of this study (see
\figref{ER}). The mismatch may cause the poor performance in the
stronger anisotropic region.  Therefore, in the following, we will
focus on the weak anisotropic region, $J_2/J_1 \ge 0.7$.
\begin{figure}
  \centering
  \includegraphics[width=0.45\textwidth]{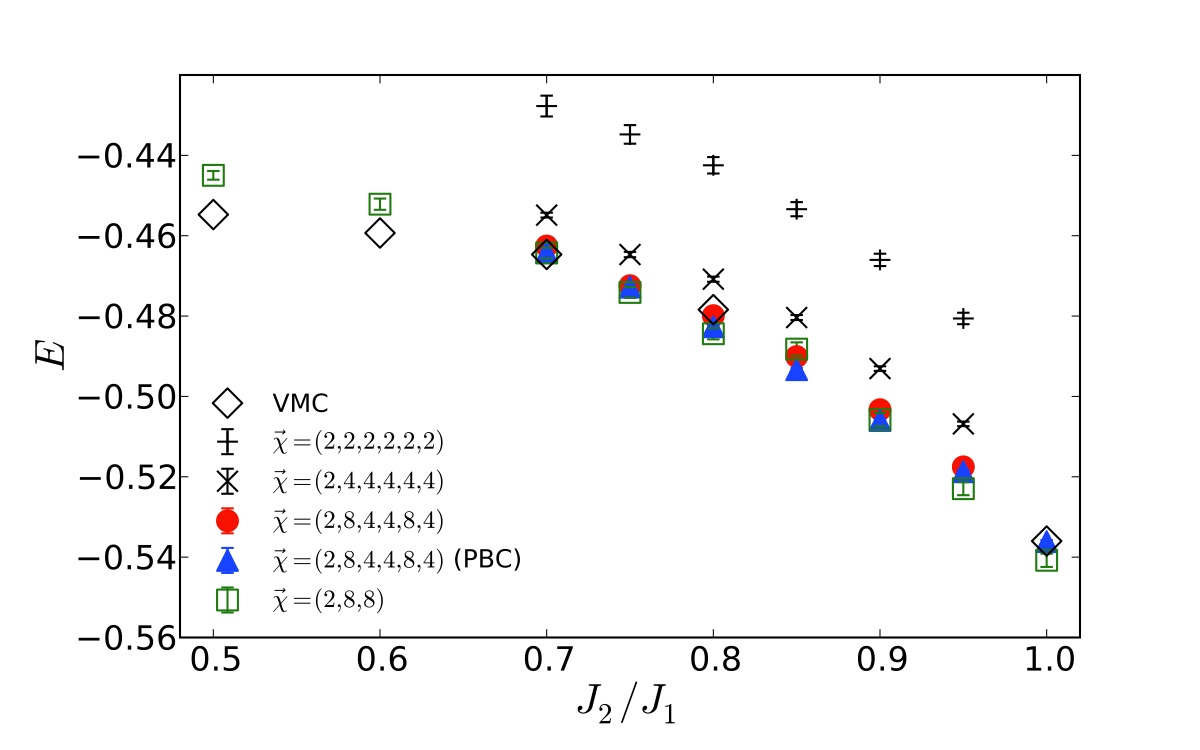}
  \caption{(Color online) Energy per site by using tensor networks with two
    ER levels ($N=2166$). The results of the infinite-size scheme are mainly
    plotted. The results of the PBC scheme with two ER levels ($N=2166$) and
    the infinite-size scheme with a single ER level ($N=114$) are also
    plotted only for the largest tensor size.  The VMC results are
    adapted from Tables I and III in Ref.~\onlinecite{Yunoki:2006hn}.}
  \figlabel{ene_two_step_ER}
\end{figure}
\begin{figure}
  \centering
  \includegraphics[width=0.45\textwidth]{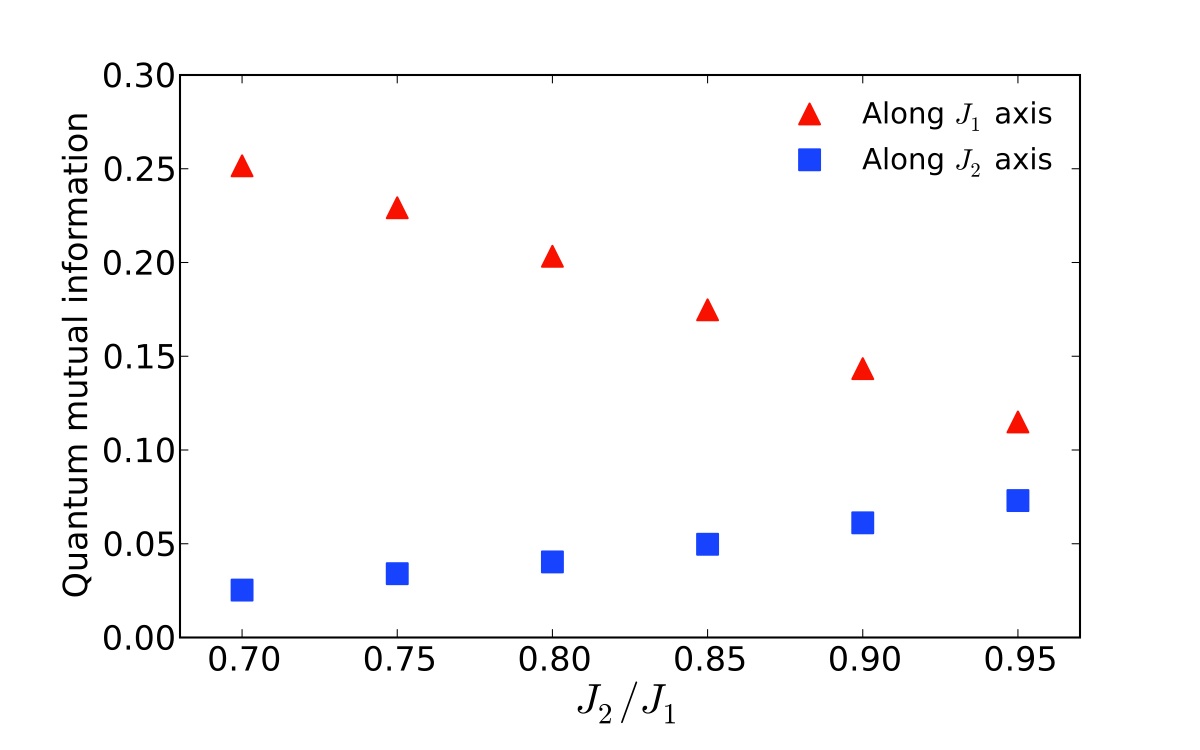}
  \caption{(Color online) Quantum mutual information of nearest
    neighbor sites along $J_1$ and $J_2$ axes using the infinite-size
    scheme with two ER levels ($N=2166$). The size of the tensors is
    $\vec{\chi} = (2,8,4,4,8,4)$.}
  \figlabel{ent_aniso}
\end{figure}
\subsubsection{Magnetization}
\begin{figure}
  \centering
  \includegraphics[width=0.45\textwidth]{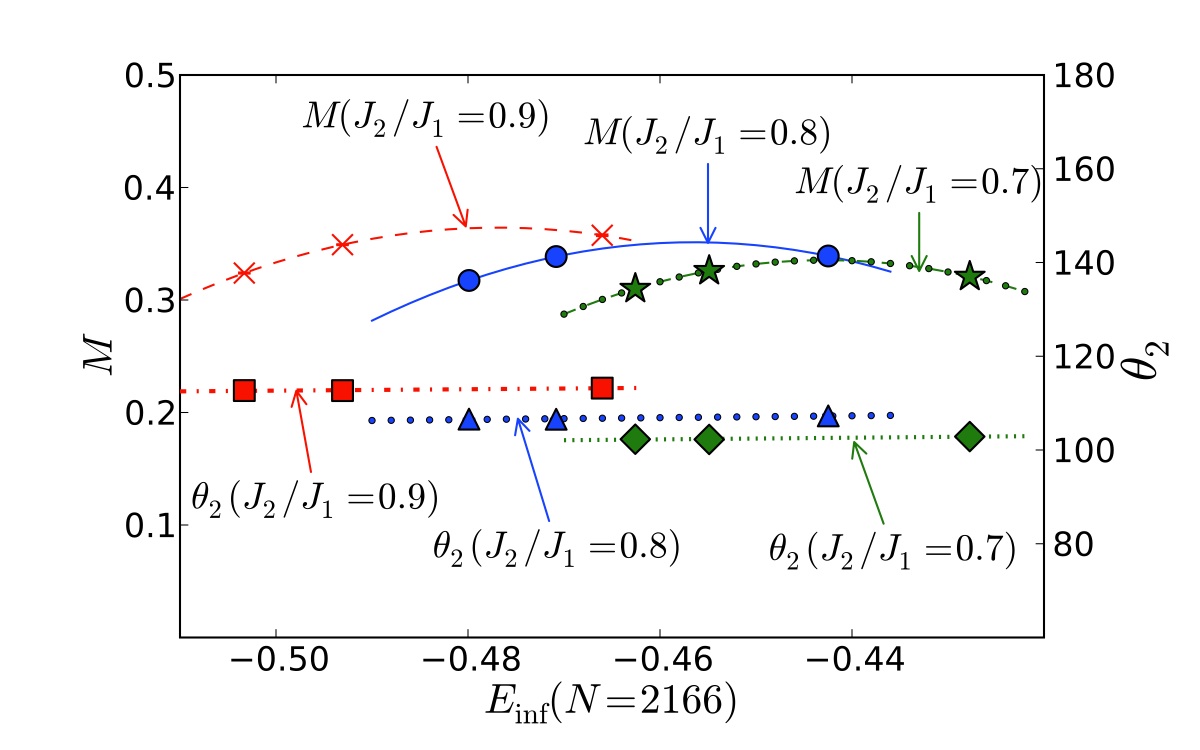}
  \caption{(Color online) Average on-site magnetization and average
    angle between magnetic moments on $J_2$ links. These results are
    obtained using the infinite-size scheme with two ER levels
    ($N=2166$). The horizontal axis denotes energy per site, $E_{\rm
      inf} (N=2166)$. The left and right vertical axes denote
    magnetization, $M$, and angle between magnetic moments on $J_2$
    links, $\theta_2$, respectively.  Crosses, circles, and stars
    denote $(E_{\rm inf} (N=2166), M)$ at $J_2/J_1 = 0.9$, $0.8$, and
    $0.7$, respectively.  Squares, triangle, and diamonds denote
    $(E_{\rm inf} (N=2166), \theta_2)$ at $J_2/J_1 = 0.9, 0.8,$ and
    $0.7$, respectively.  The three curves are quadratic fittings for
    $(E_{\rm inf} (N=2166), M)$ points.  The three lines are linear
    fittings for $(E_{\rm inf} (N=2166), \theta_2)$ points.  For all
    cases of $J_2/J_1$, the tensor sizes of the left-most, middle, and
    right-most points are $(\chi_1, \chi_2, \chi_3, \chi_4, \chi_5,
    \chi_6)= (2,8,4,4,8,4)$, $(2,4,4,4,4,4)$, and $(2,2,2,2,2,2)$,
    respectively.}
  \figlabel{ene_mag_N2166}
\end{figure}
In the region $0.7 \le J_2/J_1 \le 1$, the MERA tensor network states
break SU(2) symmetry and they have finite on-site magnetizations for
both finite and infinite-size lattices. Figure~\ref{fig:ene_mag_N2166}
shows magnetization and the average angle between magnetic moments on
$J_2$ links for three cases, $J_2/J_1=0.7$, $0.8$, and $0.9$,
calculated using infinite-size scheme. As in the isotropic case, the
dependence on tensor size remains. We cannot simply extrapolate them
in the limit of the infinite dimension.  However, even if the
ground-state energy is about $0.1$ lower than the best results of this study, the
extrapolated values from fitting curves are finite.  Therefore, 
the results of this study suggest that the ground states are magnetic. The wave function
from the infinite-size scheme is a correct quantum state in the
thermodynamic limit. Thus, at least, the magnetic state is a good
candidate for the ground state in this model.  In recent VMC
calculations\cite{Heidarian:2009gd}, the disappearance of
magnetization for $J_2/J_1 \le 0.8$ was reported. However, in
the results of this study, the magnetizations smoothly change even in the region of
$J_2/J_1 \le 0.8$.

The angle between magnetic moments of nearest neighbor sites weakly
depends on the tensor size, as shown in \figref{ene_mag_N2166}. There
are two groups of pairs of nearest neighbor sites: One is defined on
the $J_1$ links and the other is defined on the $J_2$ links. We define
$\theta_i$ as the average angle between magnetic moments on $J_i$
links. In detail, $\theta_2$ may be split into two groups,
$\theta_{2a}$ and $\theta_{2b}$, which correspond to two directions
along the $J_2$ axis. First, in all results in the anisotropic region,
the values of the sum $\theta_1+\theta_{2a}+ \theta_{2b}$ are
$359.5(6)$. Thus, all magnetic moments are coplanar, i.e., always lie
on the same plane.  Figure~\ref{fig:mangle-aniso} shows the average
angle $\theta_2$ of MERA tensor network states. Results only
for the largest tensor size are plotted. There is no discrepancy between
$\theta_{2a}$ and $\theta_{2b}$ in the cases of two ER levels.  As
shown by the solid (red) circles and the solid (blue) triangles in
\figref{mangle-aniso}, the average angle smoothly changes from
$115.9(2)$ to $102(2)$ when $J_2/J_1$ decreases from $0.95$ to
$0.7$. Thus the wave vectors of magnetic order are incommensurate. In
contrast, as shown by the open (green) squares and the open (green)
diamonds in \figref{mangle-aniso}, the average angle suddenly changes
around $J_2/J_1=0.825$ in the case of single ER. In detail, under
$J_2/J_1 \le 0.8$, $\theta_2$ splits into $\theta_{2a}$ and
$\theta_{2b}$. In other words, the reflection symmetry along the $J_1$
axis breaks in $J_2/J_1 \le 0.8$ by using the single ER level. Similar
results have been reported in a cylindrical lattice with a width as
narrow as 6 in DMRG calculations\cite{Weichselbaum:2011hq}. The reason
for the behavior of angle in the case of single ER is the strong
finite-size effect of the unit cell in tensor networks. Thus, a large
unit cell is necessary to capture the incommensurate state. The dashed
line in \figref{mangle-aniso} is the angle between magnetic moments on
the $J_2$ links for the classical spatially anisotropic
antiferromagnetic Heisenberg model. It is different from the results
of MERA tensor networks. The solid line in \figref{mangle-aniso} shows
the results of the series expansion method \cite{Zheng:1999cu}.
Although the methods are very different, the series expansion results
and those from the present work are in good agreement with each other.
This fact gives a strong evidence for existence of a stable spiral
phase. The change of wave number is larger than that of the classical
one. Quantum fluctuations weaken the effective coupling between chains
and enhance the incommensurability.

\begin{figure}
  \centering
  \includegraphics[width=0.45\textwidth]{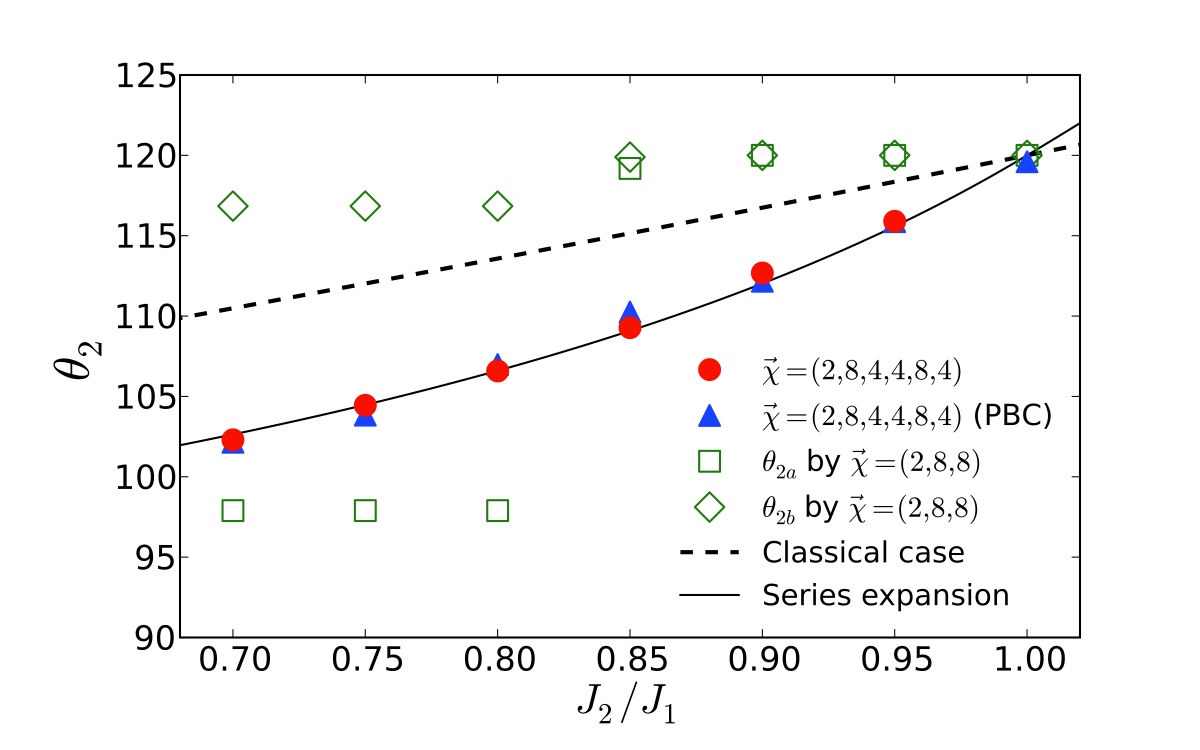}
  \caption{(Color online) Average angles between magnetic moments on
    $J_2$ links. The solid line denotes results of series expansion
    methods\cite{Zheng:1999cu}. The dotted line shows the value from
    the classical anisotropic triangular model.  }
\figlabel{mangle-aniso}
\end{figure}

\section{Conclusions}
\label{sec:conclusions}

Using ER tensor networks, we numerically studied the ground states of
spin-$\frac12$ Heisenberg antiferromagnets on anisotropic triangular
lattices.

Since the area law of entanglement entropy holds in an ER tensor
network, in principle, this new method may be effective for capturing
high entanglement quantum states as expected in frustrated quantum
magnets. Since we only assume an entanglement structure, the results of this study
will serve as a new piece of evidence independent of the previous
ones, with a totally new kind of bias.

Numerical results using tensor networks with one and two
ER levels were reported, which correspond to $N=114$ and $N=2166$ unit cells,
respectively. First, we confirmed the $120^\circ$ magnetic order
ground state at the isotropic point $J_1=J_2$ by using a tensor
network with a single ER level ($N=114$). The entanglement entropy was
more sensitive to the direct product state on the top layer with a
small unit cell size. Second, using the tensor network with two ER
levels ($N=2166$), we found a stable spiral magnetic structure with
incommensurate wave vectors at least in the anisotropic region $0.7
\le J_2/J_1 < 1$. In particular, the angle between magnetic moments on
nearest neighbor sites agrees very well with results of the series
expansion method\cite{Zheng:1999cu}, which is a very different
approach.

However, the spiral phase that we found overlaps the disordered phase
reported in VMC calculations\cite{Yunoki:2006hn,
  Heidarian:2009gd}. Although we can roughly extrapolate magnetization
in the limit of infinite dimension, we did not find the sharp decrease
in magnetization around $J_2/J_1 = 0.85$ reported in the VMC
calculation\cite{Heidarian:2009gd}.  By increasing the dimension of
tensor indices, and by modifying the structure of the ER tensor
network, the author hopes that we can obtain a complete answer in the
near future. In particular, to overcome the computational cost, we may
need to explore less demanding methods in future studies such as
combining the tensor network method with Monte Carlo
sampling\cite{Ferris:2012ko}.

In real materials such as $\kappa$-(BEDT-TTF)${}_2$ Cu${}_2$ (CN)${}_3$
and EtMe${}_3$Sb[Pd(dmit)${}_2$]${}_2$, high-order interaction may
play an import role. In particular, models with ring exchange
were discussed to explain the disordered behavior in real
materials\cite{Powell:2011ce}. The ER tensor network may be useful for
studying ground states of such models.

\begin{acknowledgments}
  The author would like to thank N.~Kawashima for stimulating
  discussions and comments on the manuscript. He also would like to
  acknowledge helpful discussions with L.~Capriotti, P.~Corboz,
  G.~Evenbly, S.~Furukawa, Y.~Kamiya, J.~Lou, R.~H.~McKenzie, M.~Sato, 
  S.~Singh, T.~Suzuki, G.~Vidal, and M.~Q.~Weng and the hospitality at
  the Kavli Institute for Theoretical Physics during the research
  program ``Disentangling quantum many-body systems: Computational and
  conceptual approaches'' supported by the National Science Foundation
  under Grant No. PHY05-51164. This research was supported in part by
  Grants-in-Aid for Scientific Research No. 22340111 and No. 23540450.
\end{acknowledgments}

\bibliographystyle{apsrev4-1}
\bibliography{tensor}

\end{document}